\RequirePackage{ifpdf}
\ifpdf % We are running pdfTeX in pdf mode
\documentclass[pdftex]{sigma}
\else
\documentclass{sigma}
\fi

\def\al{\alpha}
\def\ep{\varepsilon}
\def\l{\lambda}

\def\epsilon{\varepsilon}

\def\tilde{\widetilde}
\def\hat{\widehat}
\def\su2{{\mathfrak {su}}(2)}
\def\e3{{\mathfrak {e}}(3)}

\newcommand{\CB}{{\mathcal B}}

\newcommand{\CC}{{\mathcal C}}

\newcommand{\CR}{{\mathcal R}}
\newcommand{\CA}{{\mathcal A}}

\begin{document}

\allowdisplaybreaks

\renewcommand{\PaperNumber}{077}

\FirstPageHeading

\ShortArticleName{On Miura Transformations and Volterra-Type Equations}

\ArticleName{On Miura Transformations\\ and Volterra-Type Equations Associated\\ with the
Adler--Bobenko--Suris Equations}

\Author{Decio LEVI~$^\dag$, Matteo PETRERA~$^{\ddag\dag}$, Christian SCIMITERNA~$^{\ddag\dag}$ and Ravil YAMILOV~$^\S$ }

\AuthorNameForHeading{D. Levi, M. Petrera, C. Scimiterna and R. Yamilov}

\Address{$^\dag$~Dipartimento di Ingegneria Elettronica,Universit\`a degli Studi Roma Tre and Sezione INFN,\\
 \hphantom{$^\dag$}~Roma Tre,
Via della Vasca Navale 84, 00146 Roma, Italy}
\EmailD{\href{mailto:levi@fis.uniroma3.it}{levi@fis.uniroma3.it}}

\Address{$^\ddag$~Dipartimento di Fisica E. Amaldi,
Universit\`a degli Studi Roma Tre and Sezione INFN,\\
 \hphantom{$^\ddag$}~Roma Tre,
Via della Vasca Navale 84, 00146 Roma, Italy}
\EmailD{\href{mailto:petrera@fis.uniroma3.it}{petrera@fis.uniroma3.it}, \href{mailto:scimiterna@fis.uniroma3.it}{scimiterna@fis.uniroma3.it}}

\Address{$^\S$~Ufa Institute of Mathematics, 112
Chernyshevsky Str., Ufa 450077, Russia}
\EmailD{\href{mailto:RvlYamilov@matem.anrb.ru}{RvlYamilov@matem.anrb.ru}}

\ArticleDates{Received August 29, 2008, in f\/inal form October 30,
2008; Published online November 08, 2008}

\Abstract{We construct Miura transformations mapping
the scalar spectral problems of  the integrable lattice
equations belonging to the Adler--Bobenko--Suris (ABS) list
into the discrete Schr\"odinger spectral problem associated with Volterra-type equations. We show that the ABS equations correspond to B\"acklund transformations for some particular cases of the discrete Krichever--Novikov equation found by Yamilov (YdKN equation). This enables us to construct
new generalized symmetries for the ABS equations. The same can be said about the generalizations of the ABS equations introduced  by Tongas, Tsoubelis and Xenitidis. All of them generate B\"acklund transformations for the YdKN equation. The higher order generalized
symmetries we construct in the present paper conf\/irm their integrability.}


\Classification{37K10; 37L20; 39A05}

\section{Introduction}

The discovery of new two-dimensional integrable partial dif\/ference equations
(or $\mathbb{Z}^2$-lattice equations)
 is always a very challenging problem as, by proper continuous
limits, many other results on
dif\/ferential-dif\/ference and partial dif\/ferential equations may be
obtained. Moreover many physi\-cal and biological applications involve discrete
systems, see for instance \cite{galor,sande} and references therein.

The theory of nonlinear
integrable dif\/ferential  equations  got a boost when
Gardner, Green, Kruskal and Miura introduced the Inverse Scattering Method for the solution of the Korteweg--de~Vries equation. A summary of these
results can be found in the
Encyclopedia of Mathematical Physics \cite{emp}. A few techniques have been introduced to classify integrable partial dif\/ferential equations. Let us just
mention the classif\/ication scheme  introduced by Shabat, using the formal symmetry
approach (see \cite{msy} for a review). This approach has been successfully extended to the dif\/ferential-dif\/ference case
by Yamilov \cite{y83,y06,ly}. In the completely discrete case the situation
turns out to be quite dif\/ferent. For instance, in the case of $\mathbb{Z}^2$-lattice equations
the formal symmetry technique  does not work. In this framework, the f\/irst
exhaustive classif\/ications of families of lattice equations have been presented
in \cite{a2000} by Adler and in \cite{abs,abs1} by Adler, Bobenko and Suris.

In the present paper we shall consider
the Adler--Bobenko--Suris (ABS) classif\/ication
of $\mathbb{Z}^2$-lattice equations def\/ined on the square lattice \cite{abs}. We refer to
the papers \cite{abs1,ra,gr,lp,lps,gr1} for some recents results about these equations.
Our main purpose is the
analysis of their transformation properties.
In fact, our aim is, on the one hand,  to  present  new
Miura
transformations between the ABS equations and Volterra-type
dif\/ference equations
and on the other hand, to show that the ABS equations correspond to B\"acklund transformations for some particular cases of the discrete
Krichever--Novikov equation found by Yamilov (YdKN equation)~\cite{y83}.

Section \ref{sec1}
is devoted to a short review of the integrable $\mathbb{Z}^2$-lattice equations derived in~\cite{abs}
and  to present  details on their matrix and scalar spectral problems. In Section~\ref{sec3},
by transforming the
obtained scalar spectral problems into the discrete Schr\"odinger spectral problem
associated with the Volterra lattice we will be able
to connect the ABS
equation with
Volterra-type equations. In Section~\ref{sec4} we
prove that the ABS equations correspond to B\"acklund transformations for
certain subcases of the YdKN equation.
Using this  result and a master symmetry of the YdKN equation, we construct
new generalized symmetries for the ABS list. Then
we discuss the integrability of a~class
of non-autonomous
ABS equations and  of a generalization  of the ABS equations introduced by Tongas, Tsoubelis and Xenitidis in~\cite{gr}. Section~\ref{sec5} is devoted to some concluding remarks.

%%%%%%%%%%%%%%%%%%%%%%%%%%%%%%%
%%%%%%%%%%%%%%%%%%%%%%%%%%%%%%%
\section{A short review of the ABS equations} \label{sec1}
%%%%%%%%%%%%%%%%%%%%%%%%%%%%%%%
%%%%%%%%%%%%%%%%%%%%%%%%%%%%%%%

A two-dimensional partial dif\/ference equation is a functional relation among the values of a~function
$u: \mathbb{Z}^2 \rightarrow \mathbb C$ at dif\/ferent points of
the lattices of indices~$n$,~$m$. It involves the independent
variables $n$, $m$ and the lattice parameters $\alpha$, $\beta$
\begin{gather*}
\mathcal E (n,m, u_{n,m}, u_{n+1,m},u_{n,m+1},\dots; \alpha, \beta)=0.
\end{gather*}

For the dependent variable $u$ we shall adopt the following notation throughout  the paper
\begin{gather}
u= u_{0,0}=u_{n,m}, \qquad u_{k,l}= u_{n+k,m+l}, \qquad k,l \in \mathbb{Z}. \label{not}
\end{gather}

We  consider here the ABS list of  integrable lattice equations, namely those af\/f\/ine linear
(i.e.\ polynomial of degree one in each argument) partial dif\/ference
equations of the form
\begin{gather}
\mathcal E (u_{0,0}, u_{1,0}, u_{0,1}, u_{1,1}; \alpha, \beta) = 0, \label{jj}
\end{gather}
whose integrability is based on the {\it consistency around a cube} \cite{abs,abs1}.
The function $\mathcal E$ depends explicitly on the values of $u$ at the vertices  of an elementary quadrilateral,
i.e.\ $\partial_{u_{i,j}} \mathcal E \ne 0$, where $i,j=0,1$.
The lattice parameters $\alpha$, $\beta$ may, in general, depend on the
variables $n$, $m$, i.e.\ $\alpha = \alpha_n$, $\beta = \beta_m$.
However, we shall discuss such non-autonomous extensions  in Section~\ref{sec4}.

The complete list of the ABS equations can be found in~\cite{abs}. Their integrability holds by construction since
the consistency around a cube furnishes their Lax pairs~\cite{abs,bob,ni}.
The ABS equations are given by the list H
\begin{gather*}
{\mbox{{(H1)}}} \quad (u_{0,0}-u_{1,1})  (u_{1,0}-u_{0,1})  - \alpha  +   \beta   =  0 ,\nonumber \\
{\mbox{{(H2)}}} \quad (u_{0,0}-u_{1,1})(u_{1,0}-u_{0,1}) +(\beta-\alpha) (u_{0,0}+u_{1,0}+u_{0,1}+u_{1,1})  - \alpha^2 + \beta^2 = 0,  \nonumber \\
{\mbox{{(H3)}}}  \quad \alpha (u_{0,0} u_{1,0}+u_{0,1} u_{1,1}) - \beta (u_{0,0} u_{0,1}+u_{1,0} u_{1,1}) + \delta (\alpha^2-\beta^2) = 0,\nonumber
\end{gather*}
and the list Q
\begin{gather*}
{\mbox{{(Q1)}}}   \quad  \alpha (u_{0,0}-u_{0,1}) (u_{1,0}- u_{1,1}) - \beta (u_{0,0}- u_{1,0}) (u_{0,1} -u_{1,1})   + \delta^2 \alpha \beta (\alpha-\beta)= 0,\nonumber\\
{\mbox{{(Q2)}}}     \quad \alpha (u_{0,0}-u_{0,1}) (u_{1,0}- u_{1,1}) - \beta (u_{0,0}- u_{1,0}) (u_{0,1} -u_{1,1})  \nonumber\\
  \qquad \quad \quad{}+ \alpha \beta (\alpha-\beta) (u_{0,0}+u_{1,0}+u_{0,1}+u_{1,1}) - \alpha \beta (\alpha-\beta) (\alpha^2-\alpha \beta + \beta^2) = 0,
  \nonumber\\
{\mbox{{(Q3)}}}      \quad (\beta^2-\alpha^2) (u_{0,0} u_{1,1}+u_{1,0} u_{0,1}) + \beta (\alpha^2-1) (u_{0,0} u_{1,0}+u_{0,1} u_{1,1})\nonumber\\
  \qquad \quad \quad{} - \alpha (\beta^2-1) (u_{0,0} u_{0,1}+u_{1,0} u_{1,1}) - \frac{\delta^2 (\alpha^2-\beta^2) (\alpha^2-1) (\beta^2-1)}{4 \alpha \beta}=0 , \nonumber \\
{\mbox{{(Q4)}}}     \quad   a_0 u_{0,0} u_{1,0} u_{0,1} u_{1,1}   + a_1 (u_{0,0} u_{1,0} u_{0,1} + u_{1,0} u_{0,1} u_{1,1} + u_{0,1} u_{1,1} u_{0,0} + u_{1,1} u_{0,0} u_{1,0}) \nonumber \\
    \qquad \quad \quad{}+a_2 (u_{0,0} u_{1,1} + u_{1,0} u_{0,1}) + \bar{a}_2 (u_{0,0} u_{1,0}+u_{0,1} u_{1,1})
+ \tilde{a}_2 (u_{0,0} u_{0,1}+u_{1,0} u_{1,1})\nonumber \\
    \qquad \quad \quad{} + a_3 (u_{0,0} + u_{1,0} + u_{0,1} + u_{1,1}) + a_4 = 0.\nonumber
\end{gather*}
The coef\/f\/icients $a_i$'s appearing
 in equation~(Q4) are connected to $\alpha$ and $\beta$ by the relations
\begin{gather*}
 a_0 = a+b, \qquad a_1=-a \beta - b \alpha,\qquad a_2=a \beta^2 + b \alpha^2, \\
 \bar{a}_2 = \frac{a b (a+b)}{2 (\alpha-\beta)} + a \beta^2 - \left(2 \alpha^2 - \frac{g_2}{4}\right) b,\qquad
 \tilde{a}_2 = \frac{a b (a+b)}{2 (\beta-\alpha)} + b \alpha^2 - \left(2 \beta^2 - \frac{g_2}{4}\right) a, \\
 a_3 = \frac{g_3}{2}a_0 - \frac{g_2}{4} a_1,\qquad a_4=\frac{g_2^2}{16}a_0-g_3 a_1,
\end{gather*}
with
$a^2 = r(\alpha)$, $b^2 = r(\beta)$, $r(x)=4 x^3-g_2 x - g_3$.

Following \cite{abs} we remark that
\begin{itemize}\itemsep=0pt
\item Equations (Q1)--(Q3) and (H1)--(H3) are all degenerate subcases of equation~(Q4) \cite{atkinson}.

\item Parameter $\delta$ in equations (H3), (Q1) and (Q3) can be rescaled, so that one
can assume without loss of generality that $\delta= 0$ or $\delta=1$.

\item The original ABS list  contains two further equations (list A)
\begin{gather*}
{\mbox{{(A1)}}} \quad \alpha (u_{0,0}+u_{0,1})\, (u_{1,1}+u_{1,0})\, -\,\beta (u_{0,0}+u_{1,0})\, (u_{1,1}+u_{0,1})
  - \delta^2 \alpha \beta (\alpha-\beta)
= 0, \nonumber \\
{\mbox{{(A2)}}}   \quad (\beta^2 -\alpha^2) (u_{0,0} u_{1,0} u_{0,1} u_{1,1}+1) +\beta (\alpha^2 -1 )
(u_{0,0} u_{0,1}+u_{1,0} u_{1,1} ) \nonumber \\
 \qquad \quad \quad{}  - \alpha (\beta^2-1) ( u_{0,0} u_{1,0}+u_{0,1} u_{1,1} )= 0. \nonumber
\end{gather*}
Equations (A1) and (A2) can be transformed by an extended
group of M\"obius transformations into equations (Q1) and (Q3) respectively. Indeed, any
solution $u=u_{n,m}$ of~(A1) is transformed into a solution $\tilde u =\tilde u_{n,m}$
of (Q1) by $u_{n,m} = (-1)^{n+m} \tilde u_{n,m}$ and
any
solution $u= u_{n,m}$ of (A2) is transformed into a solution $\tilde u= \tilde u_{n,m}$
of (Q3) with $\delta = 0$ by $u_{n,m} = \left(\tilde u_{n,m}\right)^{(-1)^{n+m}}$.
\end{itemize}

Some of the above equations were known before Adler, Bobenko and Suris
presented their classif\/ication, see for instance \cite{nc,hi}.
We f\/inally recall that a more general classif\/ication of integrable lattice equations def\/ined on the square
has been recently carried out by Adler, Bobenko and Suris  in \cite{abs1}.
But here we shall consider only the lists H and Q contained in~\cite{abs}.

\subsection{Spectral problems of the ABS equations} \label{sec2}

The algorithmic  procedure described in \cite{abs,bob,ni} produces
a $2 \times 2$ matrix Lax pair for the ABS equations, thus ensuring their integrability. It may be written as
\begin{gather}
 \Psi_{1,0}=L(u_{0,0},u_{1,0}; \alpha, \lambda)   \Psi_{0,0} , \qquad \Psi_{0,1}=M(u_{0,0},u_{0,1}; \beta, \lambda) \Psi_{0,0}, \label{gh}
\end{gather}
with $\Psi=(\psi(\l) ,\phi(\l) )^T$,
where the lattice parameter $\lambda$ plays the role of the spectral parameter. We shall use the following
notation
\begin{gather*}
 L(u_{0,0},u_{1,0}; \alpha, \lambda)= \frac{1} {\ell}\left(\begin{array} {cc}
L_{11} & L_{12}\\
L_{21} & L_{22}\end{array}\right), \qquad
M(u_{0,0},u_{0,1}; \beta, \lambda)=\frac{1} {t}\left(\begin{array} {cc}
M_{11} & M_{12}\\
M_{21} & M_{22}\end{array}\right),
\end{gather*}
where $\ell=\ell_{0,0}= \ell (u_{0,0},u_{1,0}; \alpha, \lambda)$, $t= t_{0,0}= t(u_{0,0},u_{0,1}; \beta, \lambda)$,
$L_{ij}=L_{ij}(u_{0,0},u_{1,0}; \alpha, \lambda)$ and
$M_{ij}=M_{ij}(u_{0,0},u_{0,1}; \beta, \lambda)$, $i,j=1,2$.
The matrix $M$ can be
obtained from $L$ by replacing $\alpha$ with $\beta$ and shifting along direction~2 instead of~1.
In Table~\ref{tab1} we give the entries of the matrix~$L$ for the ABS equations.

\begin{table}[t]\footnotesize
%\begin{sidewaystable}[h!]
\centering\caption{Matrix $L$ for the ABS equations
(in equation~(Q4) $a^2 = r(\alpha)$, $b^2 = r(\l)$, $r(x)=4 x^3-g_2 x - g_3$).} \label{tab1}

\vspace{1mm}

\begin{tabular}{||@{\,\,}l@{\,\,}||@{\,}c@{\,}|@{\,}c@{\,}|@{\,}c@{\,}|@{\,}c@{\,}||}
 \hline
\bsep{1ex}\tsep{1ex} & $L_{11}$ & $L_{12}$ & $L_{21}$ & $L_{22}$\\
\hline \hline
\bsep{1ex}\tsep{1ex} {\mbox{{H1}}}& $u_{0,0}-u_{1,0}$ & $(u_{0,0}-u_{1,0})^2+\alpha-\lambda$ & $1$ & $u_{0,0}-u_{1,0}$\\
 \hline
\tsep{1ex}
{\mbox{{H2}}} & $u_{0,0}-u_{1,0}+\alpha-\lambda$ & $(u_{0,0}-u_{1,0})^2+2(\alpha-\lambda)(u_{0,0}+u_{1,0})+$ & $1$ & $u_{0,0}-u_{1,0}-\alpha+\lambda$\\
\bsep{1ex} & & $+\alpha^2-\lambda^2$& & \\
 \hline
\bsep{1ex}\tsep{1ex}
{\mbox{{H3}}}& $\lambda u_{0,0}-\alpha u_{1,0}$ & $\lambda(u_{0,0}^2+u_{1,0}^2)-2\alpha u_{0,0}u_{1,0}+\delta(\lambda^2-\alpha^2)$ & $\alpha$ & $\alpha u_{0,0}-\lambda u_{1,0}$\\
 \hline
\bsep{1ex}\tsep{1ex}
{\mbox{{Q1}}}& $\lambda(u_{1,0}-u_{0,0})$ & $-\lambda(u_{1,0}-u_{0,0})^2+\delta\alpha\lambda(\alpha-\lambda)$ & $-\alpha$ & $\lambda(u_{1,0}-u_{0,0})$\\
 \hline
\tsep{1ex}
{\mbox{{Q2}}}& $\lambda(u_{1,0}-u_{0,0})+$ & $-\lambda(u_{1,0}-u_{0,0})^2+$ & $-\alpha$ & $\lambda(u_{1,0}-u_{0,0})-$\\
 & $+\alpha\lambda(\alpha-\lambda)$& $+2\alpha\lambda(\alpha-\lambda)(u_{1,0}+u_{0,0})-$ & & $-\alpha\lambda(\alpha-\lambda)$\\
\bsep{1ex} &&$-\alpha\lambda(\alpha-\lambda)(\alpha^2-\alpha\lambda+\lambda^2)$&&\\
\hline
\tsep{1ex}
{\mbox{{Q3}}}& $\alpha(\lambda^2-1)u_{0,0}-$ & $-\lambda(\alpha^2-1)u_{0,0}u_{1,0}+$ & $\lambda(\alpha^2-1)$ & $(\lambda^2-\alpha^2)u_{0,0}-$\\
\bsep{1ex} & $-(\lambda^2-\alpha^2)u_{1,0}$& $+\delta(\alpha^2-\lambda^2)(\alpha^2-1)(\lambda^2-1)/(4\alpha\lambda)$ & & $-\alpha(\lambda^2-1)u_{1,0}$\\
 \hline
\tsep{1ex}
 {\mbox{{Q4}}}& $-a_1 u_{0,0}u_{1,0}-$
 & $-\bar{a}_2 u_{0,0}u_{1,0}-a_3(u_{0,0}+u_{1,0})-a_4$ & $a_0 u_{0,0}u_{1,0} +\bar{a}_2+$ & $a_1 u_{0,0}u_{1,0}+a_2 u_{0,0}+$\\
\bsep{1ex} & $- a_2 u_{1,0}-\tilde{a}_2 u_{0,0}-a_3$ & & $+a_1(u_{0,0}+u_{1,0})$ & $+\tilde{a}_2 u_{1,0}+a_3$\\
 \hline
\end{tabular}
\end{table}

Note that
$\ell$ and $t$ are computed by requiring that the compatibility condition
between~$L$ and~$M$
produces the ABS equations
 (H1)--(H3) and (Q1)--(Q4).
The factor $\ell$ can be written as
\begin{gather}
\ell_{0,0} = f (\alpha,\lambda) [\rho(u_{0,0},u_{1,0}; \alpha)]^{1/2}, \label{bv}
\end{gather}
where the functions $f=f (\alpha,\lambda)$ is an arbitrary normalization
factor. The functions $f=f (\alpha,\lambda)$
and
$\rho= \rho_{0,0}=\rho(u_{0,0},u_{1,0}; \alpha)$ for equations~(H1)--(H3) and (Q1)--(Q4)
are given in Table~\ref{tab2}.
A~formula similar to (\ref{bv}) holds also for the factor $t$.

The scalar Lax pairs for the ABS equations may be immediately computed from equation~(\ref{gh}). Let us write
 the scalar equation  just for the second component $\phi$ of the vector $\Psi$ (the use of the f\/irst component would give similar results).
For
equations~(H1)--(H3) and (Q1)--(Q3) it reads
\begin{gather}
(\rho_{1,0})^{1/2} \phi_{2,0}- (u_{2,0}-u_{0,0}) \phi_{1,0}+ (\rho_{0,0})^{1/2} \mu   \phi_{0,0}=0, \label{spo}
\end{gather}
where the explicit expressions of
$\mu=\mu(\alpha,\lambda)$ are given in Table~\ref{tab2}.
The corresponding scalar equation for equation~(Q4) takes a dif\/ferent form and needs a separate analysis which will be done in a separate work.

\begin{table}[t]\footnotesize
\centering\caption{Scalar spectral problems for the ABS equations
(in equation~(Q4) $c^2=r(\lambda)$, $r(x)=4 x^3-g_2 x - g_3$).  }\label{tab2}
\vspace{1mm}
\begin{tabular}{||@{\,\,}l@{\,\,}||@{\,}c@{\,}|@{\,}c@{\,}||@{\,}c@{\,}||} \hline
\bsep{1ex}\tsep{1ex} & $f (\alpha,\lambda) $ & $\rho(u_{0,0},u_{1,0}; \alpha)$ &
 $\mu(\alpha,\lambda)$ \\
\hline \hline
\bsep{1ex}\tsep{1ex}
 {\mbox{{H1}}} & $-1$ & $1$ & $\lambda-\alpha$ \\
\hline
\bsep{1ex}\tsep{1ex}  {\mbox{{H2}}}&   $-1$ & $u_{0,0}+u_{1,0}+\alpha$ & $2(\lambda-\alpha)$ \\
\hline
\bsep{1ex}\tsep{2ex}
 {\mbox{{H3}}}& $-\lambda$ &$u_{0,0}u_{1,0}+\delta\alpha$ & ${\displaystyle{ \frac{\al^2 - \lambda^2}{ \al \l^2}}}$\\
 \hline
\bsep{1ex}\tsep{2ex}
 {\mbox{{Q1}}}&   $\lambda$ & $(u_{1,0}-u_{0,0})^2- \delta^2 \alpha^{2}$ & ${\displaystyle{ \frac{\lambda - \alpha}{ \l}}}$ \\
\hline
\bsep{1ex}\tsep{2ex}
 {\mbox{{Q2}}}& $\lambda$ & $(u_{1,0}-u_{0,0})^2-2\alpha^2(u_{1,0}+u_{0,0})+\alpha^4$ & ${\displaystyle{ \frac{\lambda - \alpha}{ \l}}}$ \\
 \hline
\bsep{1ex}\tsep{2ex}
 {\mbox{{Q3}}}& $\alpha(1-\lambda^2)$ &
 $\alpha(u_{0,0}^2+u_{1,0}^2)-(\alpha^2+1)u_{0,0}u_{1,0}+
{\displaystyle{\frac{\delta^2(\alpha^2-1)^2}{4\alpha}}}$ &
 ${\displaystyle{ \frac{\al^2 - \lambda^2}{ \al^2 (1 - \l^2) }}}$   \\
\hline
\tsep{1ex}
  {\mbox{{Q4}}}&  $ (\alpha -\lambda) c^{1/2} \times$ &
  $(u_{0,0}u_{1,0}+\alpha u_{0,0}+\alpha u_{1,0}+g_2/4)^2-$ & $-$
  \\
\bsep{2ex}  & $ \times  \left[2a +  c + \frac14 \left(  \frac{a+c}{\alpha -\lambda}\right)^3 -\frac{3 \alpha (a +c )}{\alpha -\lambda} \right]^{1/2}$ &
  $- (u_{0,0}+u_{1,0}+\alpha)(4\alpha u_{0,0}u_{1,0}-g_3)$& \\
 \hline
\end{tabular}
\end{table}

\section[Miura transformations for equations (H1)-(H3) and (Q1)-(Q3)]{Miura transformations for equations (H1)--(H3)\\ and (Q1)--(Q3)} \label{sec3}

The aim of this Section is to show the existence of  a Miura transformation mapping the scalar spectral problem (\ref{spo})
of equations (H1)--(H3) and (Q1)--(Q3) into the discrete Schr\"odinger spectral problem
associated with the Volterra lattice \cite{ck}
\begin{gather}
\phi_{-1,0}+ v_{0,0}   \phi_{1,0} =  p(\lambda )  \phi_{0,0},\label{prattica}
\end{gather}
where $v_{0,0}$ is the potential of the spectral problem and the function $p(\lambda)$ plays the role
of the spectral parameter.

Suppose that a function $s_{0,0}=s(u_{0,0}, u_{1,0},
u_{0,1},\dots)$ is given by the linear equation
\begin{gather}
\frac{s_{0,0}}{s_{1,0}}= \frac{u_{2,0}-u_{0,0}}{(\rho_{0,0})^{1/2}} . \label{ggg}
\end{gather}
By performing the transformation
$
\phi_{0,0} \mapsto \mu^{n/2} s_{0,0}  \phi_{0,0},
$
and taking into account equation (\ref{ggg}),
equation~(\ref{spo}) is mapped into the scalar spectral problem (\ref{prattica})
with
\begin{gather}
v_{0,0} = \frac{ \rho_{0,0}}{(u_{1,0}- u_{-1,0})(u_{2,0}-u_{0,0})}, \qquad
p(\l)= [\mu(\alpha,\lambda)]^{-1/2}. \label{pdio}
\end{gather}
From these results there follow some remarkable consequences: (i)~There exists a Miura transformation between all equations of the set (H1)--(H3) and  (Q1)--(Q3).
Some results on this claim can be found in \cite{atkinson}; (ii)~The Miura transformation~(\ref{pdio}) can be inverted by solving a linear dif\/ference equation. Therefore we can in principle
use these remarks to f\/ind explicit solutions of the ABS equations in terms of the solutions of the Volterra equation.

The following statement holds.

\begin{proposition} \label{pp1}
The field $u$  for equations~{\rm (H1)--(H3)} and {\rm(Q1)--(Q3)} can be expressed in terms of  the
potential $v$ of the spectral problem~\eqref{prattica}
through the solution of the following linear difference equations
\begin{gather}
 {\rm{{H1:}}} \quad  u_{2,0}-
\frac{(v_{0,0}+v_{-1,0})}{v_{0,0}}u_{0,0}+\frac{v_{-1,0}}{v_{0,0}}u_{-2,0}=0, \label{r1} \\
 {\rm{{H2:}}} \quad  u_{2,0}-\frac{v_{0,0}+v_{-1,0}}{v_{0,0}}u_{0,0}+
\frac{v_{-1,0}}{v_{0,0}}u_{-2,0}-\frac{1}{v_{0,0}}=0,\label{r2} \\
 {\rm{{H3:}}} \quad u_{2,0}-\frac{1+v_{0,0}+v_{-1,0}}{v_{0,0}}u_{0,0}+
\frac{v_{-1,0}}{v_{0,0}}u_{-2,0}=0,\label{r3}\\
 {\rm{{Q1:}}} \quad
u_{2,0}-\frac{1}{v_{0,0}}u_{1,0}+\frac{2-v_{0,0}-v_{-1,0}}{v_{0,0}}u_{0,0}-
\frac{1}{v_{0,0}}u_{-1,0}+\frac{v_{-1,0}}{v_{0,0}}u_{-2,0}=0,\label{r4} \\
 {\rm{{Q2:}}} \quad u_{2,0}-\frac{1}{v_{0,0}}u_{1,0}+
\frac{2-v_{0,0}-v_{-1,0}}{v_{0,0}}u_{0,0}-
\frac{1}{v_{0,0}}u_{-1,0}+\frac{v_{-1,0}}{v_{0,0}}u_{-2,0}+
\frac{2\alpha^2}{v_{0,0}}=0,  \label{r5} \\
 {\rm{{Q3:}}} \quad  u_{2,0}-\frac{\alpha}{v_{0,0}} u_{1,0}
+\frac{\alpha^2+1-
v_{0,0}-v_{-1,0}}{v_{0,0}}u_{0,0}-\frac{\alpha}{v_{0,0}} u_{-1,0}+\frac{v_{-1,0}}{v_{0,0}}u_{-2,0}=0. \label{r6}
\end{gather}
\end{proposition}

\begin{proof} From equation~(\ref{pdio}) we get
\begin{gather}
v_{0,0}(u_{2,0} - u_{0,0}) = \frac{\rho_{0,0} }{u_{1,0} - u_{-1,0}}, \qquad
v_{-1,0}(u_{0,0} - u_{-2,0}) = \frac{\rho_{-1,0}}{u_{1,0} - u_{-1,0}}.
\nonumber \end{gather}
Subtracting these relations and taking into account that (see equation~(A.11) in \cite{gr})
\begin{gather*}
\partial_{u_{1,0}} \rho_{0,0}+ \partial_{u_{-1,0}} \rho_{-1,0}
= 2  \frac{\rho_{0,0} - \rho_{-1,0}}{u_{1,0} - u_{-1,0}}, %\label{greece}
\end{gather*}
one arrives at
\begin{gather}
v_{0,0}(u_{2,0} - u_{0,0})
- v_{-1,0}(u_{0,0} - u_{-2,0}) = \frac12 \left(\partial_{u_{1,0}} \rho_{0,0}+ \partial_{u_{-1,0}} \rho_{-1,0}  \right).
\label{greece2}
\end{gather}

Writing equation~(\ref{greece2}) explicitly for equations~(H1)--(H3) and (Q1)--(Q3) we obtain equations~(\ref{r1})--(\ref{r6}).
\end{proof}

%%%%%%%%%%%%%%%%%%%%%%%%%%%%%%%
%%%%%%%%%%%%%%%%%%%%%%%%%%%%%%%
\section{Generalized symmetries of the ABS equations} \label{sec4}
%%%%%%%%%%%%%%%%%%%%%%%%%%%%%%%
%%%%%%%%%%%%%%%%%%%%%%%%%%%%%%%

Lie symmetries of equation~(\ref{jj}) are given by those continuous
transformations which leave the equation invariant. We refer to \cite{lw6,y06} for a review on symmetries
of discrete equations.

From the inf\/initesimal point of view,
Lie symmetries
are obtained by requiring the inf\/i\-ni\-te\-si\-mal invariant condition
\begin{gather} \label{cca2}
\big(  {\rm pr} \, \widehat X_{0,0} \big) \mathcal{E}   \big|_{\mathcal{E} =0} =0,
\end{gather}
where
\begin{gather} \label{ccb2}
 \widehat X_{0,0} = F_{0,0} ( u_{0,0}, u_{1,0}, u_{0, 1}, \ldots) \partial_{u_{0,0}}.
 \end{gather}
By $ {\rm pr} \, \widehat X_{0,0}$ we mean the prolongation of the inf\/initesimal generator
$\widehat X_{0,0}$  to all points appearing in $\mathcal{E}=0$.

If $F_{0,0} = F_{0,0}( u_{0,0})$ then we get {\it point symmetries} and the procedure to construct them from equation~(\ref{cca2})
is purely algorithmic \cite{lw6}. If $F_{0,0} = F_{0,0} ( u_{0,0}, u_{1,0}, u_{0,1}, \ldots)$ the obtained symmetries are called
{\it generalized symmetries}.
In the case of nonlinear discrete equations,
the Lie point symmetries are not very common, but, if the equation is integrable, it is possible to construct
an inf\/inite family of generalized symmetries.

In correspondence with the inf\/initesimal generator (\ref{ccb2})
we can in principle construct
a group transformation by integrating the initial boundary problem
\begin{gather} \label{s1}
\frac{d  u_{0,0}(\ep)}{d \ep} = F_{0,0} (  u_{0,0}(\ep),
 u_{1,0}(\ep),  u_{0,1}(\ep), \ldots),
\end{gather}
with $ u_{0,0}(\ep =0) = v_{0,0}$, where   $\ep \in \mathbb{R}$ is the
continuous Lie group parameter and $v_{0,0}$ is a solution of equation~(\ref{jj}).
This can be done ef\/fectively only in the case of point symmetries as in the
generalized
case we have a nonlinear dif\/ferential-dif\/ference equation for which we cannot f\/ind the general
solution , but, at most, we can construct particular
solutions.

Equation~(\ref{cca2}) is equivalent to the request that the $\ep$-derivative of the equation $\mathcal{E}=0$, written for $ u_{0,0}(\ep)$, is identically satisf\/ied on  its solutions when the $\ep$-evolution of $ u_{0,0}(\ep)$ is given by equation~(\ref{s1}). This is also equivalent to say  that the f\/lows (in the group parameter space) given by equation~(\ref{s1}) are compatible or commute with $\mathcal{E}=0$.

In the papers~\cite{ra,gr} the three and f\/ive-point generalized symmetries have
been found for all equations of the ABS list. We shall use these results
 to show that the ABS equations may be interpreted as
B\"acklund transformations for the dif\/ferential-dif\/ference YdKN equation
\cite{y83}. This observation will allow us to provide an inf\/inite class of
generalized symmetries for the lattice equations belonging to the ABS list. We
shall also discuss the non-autonomous case and the generalizations of the ABS equations considered in~\cite{gr}.

\subsection{The ABS equations as B\"acklund transformations of the YdKN
equation}

In the following we show that the ABS equations may be seen as B\"acklund transformations
of the YdKN equation. Moreover we prove that the symmetries of the ABS equations
\cite{ra,gr} are
subcases of the
YdKN equation. For the sake of clarity we
consider in a more detailed way just the case of equation~(H3). Similar results can be
obtained for the whole ABS list (see Proposition~\ref{pp2}).

According to \cite{ra,gr} equation~(H3) admits the compatible three-point generalized symmetries
\begin{gather}
 \frac{d u_{0,0}}{d \ep} = \frac {u_{0,0} (u_{1,0} + u_{-1,0}) + 2 \alpha \delta}{u_{1,0} - u_{-1,0}}, \label{h3s1} \\
 \frac{d u_{0,0}}{d \ep}= \frac {u_{0,0}  (u_{0,1} + u_{0,-1}) + 2 \beta \delta}{u_{0,1} - u_{0,-1}} .  \label{h3s2}
\end{gather}
Notice that under the discrete map $n \leftrightarrow
m$, $\alpha \leftrightarrow \beta$, equation~(\ref{h3s1}) goes into
equation~(\ref{h3s2}), while equation~(H3) is left invariant.

The compatibility between equation~(H3) and equation~(\ref{h3s1}) generates a B\"acklund transformation (see an explanation below) of any solution $u_{0,0}$ of equation~(\ref{h3s1}) into its new solution
\begin{gather}
\tilde u_{0,0} = u_{0,1}, \qquad \tilde u_{1,0} = u_{1,1}.  \label{bt1}
\end{gather}
Thus equation~(H3) can be rewritten as a B\"acklund transformation for the dif\/ferential-dif\/ference equation (\ref{h3s1})
\begin{gather}
\alpha (u_{0,0} u_{1,0} + \tilde u_{0,0} \tilde u_{1,0}) - \beta
(u_{0,0} \tilde u_{0,0} + u_{1,0} \tilde u_{1,0}) + \delta (\alpha^2 -
\beta^2) = 0 .  \label{bt}
\end{gather}
Moreover, the discrete symmetry $n \leftrightarrow m$, $\alpha \leftrightarrow \beta$ implies the existence of the B\"acklund transformation for equation~(\ref{h3s2})
\begin{gather*}
\hat u_{0,0} = u_{1,0}, \qquad \hat u_{0,1} = u_{1,1} .  %\label{bt2}
\end{gather*}

This interpretation of lattice equations as B\"acklund transformations has been
discussed for the f\/irst time in the dif\/ferential-dif\/ference case in~\cite{l}. Examples of B\"acklund transformations similar to equation~(\ref{bt}) for
Volterra-type equations can be found in \cite{y94,cl}.

In \cite{ra,gr} generalized symmetries have been obtained for autonomous ABS
equations, i.e.\ such that $\alpha$, $\beta$ are constants. We  present here
some results on the non-autonomous case when~$\alpha$ and~$\beta$ depend on~$n$ and~$m$. Similar results can be found in~\cite{ra}.

Let the lattice parameters in equation~(\ref{jj}) be such that $\alpha$ is a
constant and $\beta = \beta_0 = \beta_m$. Let us consider the following two forms of equation~(\ref{jj})
\begin{gather}\label{fo1}
u_{1,1} = \xi (u_{0,0}, u_{1,0}, u_{0,1}; \alpha, \beta_0),
\qquad
u_{0,1} = \zeta (u_{0,0}, u_{1,0}, u_{1,1}; \alpha, \beta_0),  %\label{fo2}
\end{gather}
and a symmetry
\begin{gather}
 \frac{d u_{0,0}}{d \ep} = f_{0,0} = f(u_{1,0}, u_{0,0}, u_{-1,0}; \alpha),  \label{fs}
\end{gather}
given by equation~(\ref{h3s1}). We suppose that $u_{k,l}$ depends on $\ep$ in all equations and
 write down the compatibility condition between  equation~(\ref{fo1}) and equation~(\ref{fs})
\begin{gather}
f_{1,1} =  f_{0,0} \partial_{u_{0,0}} \xi  + f_{1,0}  \partial_{u_{1,0}} \xi +  f_{0,1}  \partial_{u_{0,1}}\xi .  \label{cc}
\end{gather}
As a consequence of equations~(\ref{fo1}), (\ref{fs}) the functions
$f_{1,1}$, $f_{1,0}$ and $f_{0,1}$ may be expressed in terms of the f\/ields~$u_{k,0}$,~$u_{0,l}$. Therefore, equation~(\ref{cc}) depends explicitly only on the variab\-les~$u_{k,0}$,~$u_{0,l}$,
which can be considered here as independent variables for any f\/ixed~$n$,~$m$. For
all autonomous ABS equations, the compatibility condition (\ref{cc}) is
satisf\/ied identically for all values of these variables and of the constant
parameter $\beta$. In the non-autonomous case, equation~(\ref{cc}) depends only  on
$\beta_0$ and $\alpha$.
Therefore the compatibility condition is satisf\/ied also for any $m$.

So, equation~(\ref{h3s1}) is compatible with equation~(H3) also in the case when $\alpha$ is constant, but $\beta = \beta_m$.
In a similar way, one can prove that  equation~(\ref{h3s2}) is the generalized symmetry of equation~(H3) if $\beta$ is constant, but $\alpha = \alpha_n$.

Let us now discuss the  interpretation of the ABS equations as B\"acklund transformations. Let $u_{0,0}$ be a solution of equation~(\ref{fs}), and the function $\tilde u_{0,0} = \tilde u_{n,m}(\ep)$ given by equation~(\ref{bt1}) be a solution of equation~(\ref{fo1}), which is compatible with equation~(\ref{fs}). equation~(\ref{fo1}) can be rewritten as the ordinary dif\/ference equation
\begin{gather}
 \tilde u_{1,0} = \xi (u_{0,0}, u_{1,0}, \tilde u_{0,0}; \alpha, \beta_0),  \label{btr}
 \end{gather}
where $\alpha$ is  constant, $\beta_0 = \beta_m$, $m$ is f\/ixed, $n \in \mathbb{Z}$. Dif\/ferentiating equation~(\ref{btr}) with respect to~$\ep$ and using equation~(\ref{fs}) together with the compatibility condition (\ref{cc}), one gets
\[
\frac{d \tilde  u_{1,0}}{d \ep} - \frac{d \tilde u_{0,0}}{d \ep}  \partial_{\tilde u_{0,0}}  \xi  =
f_{0,0}  \partial_{u_{0,0}}  \xi +   f_{1,0} \partial_{u_{1,0}} \xi = \tilde f_{1,0} - \tilde f_{0,0}  \partial _{\tilde u_{0,0}} \xi   ,
 \]
where
 \[
\tilde f_{k,0} =  f(\tilde u_{k+1,0}, \tilde u_{k,0}, \tilde u_{k-1,0}; \alpha) = f_{k,1} ,
\qquad \tilde u_{k,0} = u_{k,1} .
 \]
The resulting  equation is expressed in the form
\begin{gather}
\Xi_{1,0} =   \Xi_{0,0} \partial_{\tilde u_{0,0}} \xi , \qquad
\Xi_{k,0} = \frac{d \tilde u_{k,0}}{d \ep} - \tilde f_{k,0} .  \label{PhEq}
 \end{gather}

There is for the ABS  equations a formal condition $\partial_{\tilde u_{0,0}}  \xi= \partial_{u_{0,1}} \xi \ne 0$. We suppose here that, for the functions $u_{0,0}$, $\tilde u_{0,0}$ under consideration, $\partial_{\tilde u_{0,0}}  \xi \ne 0$ for all $n \in \mathbb Z$.
The function~$\tilde u_{0,0}$ is def\/ined by equation~(\ref{btr}) up to an integration function $\mu_0 = \mu_m(\ep)$. We require that $\mu_0$ satisf\/ies the f\/irst order ordinary dif\/ferential equation given by $\Xi_{0,0}|_{n=0} = 0$. Then equation~(\ref{PhEq}) implies that $\Xi_{0,0} = 0$ for all $n$, i.e.~$\tilde u_{0,0}$ is a solution of equation~(\ref{fs}).

So, we start with a  solution of a generalized symmetry of the form
(\ref{fs}), def\/ine a function $\tilde u_{0,0}$ by the dif\/ference equation
(\ref{btr}) which is a form of corresponding ABS equation, then we specify the
integration function $\mu_0$ by the ordinary dif\/ferential equation
$\Xi_{0,0}|_{n=0} = 0$, and thus obtain a new solution of equation~(\ref{fs}). This
solution depends on an integration constant $\nu_0 = \nu_m$ and the parameter
$\beta_0$. We can construct in this way the solutions $u_{0,2}, u_{0,3},
\dots, u_{0,N}$, and the last of them will depend on $2N$ arbitrary constants
$\nu_0, \beta_0, \nu_1, \beta_1, \dots, \nu_{N-1}, \beta_{N-1}$. Using such
B\"acklund transformation and starting with a simple initial solution, one can
obtain, in principle, a multi-soliton solution. See \cite{AV,atk2} for the
construction of some examples of solutions.

The symmetries (\ref{h3s1}), (\ref{h3s2}) are  Volterra-type equations, namely
\begin{gather}
 \frac{d u_{0}}{d \ep} =f(u_{1}, u_0, u_{-1}),  \label{vtype}
\end{gather}
where we have dropped one of the independent indexes $n$ or $m$, since it does not vary.
The Volterra equation corresponds to $f(u_{1}, u_0, u_{-1})= u_0 (u_{1} - u_{-1})$. An exhaustive list of dif\/ferential-dif\/ference integrable equations of the form (\ref{vtype}) has been obtained in~\cite{y83} (details can be found in~\cite{y06}). All three-point generalized symmetries of the ABS equations, with no explicit dependence on~$n$,~$m$, have the same structure as equation~(\ref{h3s1}) (see details in Section~\ref{sec5last} below) and are particular cases of the YdKN equation
\begin{gather}
 \frac{d u_{0}}{d \ep}= \frac {R(u_{1}, u_0, u_{-1})}{u_1 - u_{-1}} ,
 \qquad R(u_{1}, u_0, u_{-1})= R_0=A_0 u_1 u_{-1} + B_0 (u_1 + u_{-1}) + C_0 ,
 \label{Req}
\end{gather}
where
\begin{gather*}
 A_0 = c_1 u_0^2 + 2 c_2 u_0 + c_3, \qquad
 B_0 = c_2 u_0^2 + c_4 u_0 + c_5 ,  \qquad
 C_0 = c_3 u_0^2 + 2 c_5 u_0 + c_6 ,
\end{gather*}
and the $c_i$'s are constants. equation~(\ref{Req}) has been found by Yamilov in \cite{y83},
discussed in \cite{msy,asy}, and in most detailed form in~\cite{y06}. Its continuous limit
goes into the Krichever--Novikov equation~\cite{kn}. This is the only integrable example of the form (\ref{vtype}) which cannot be reduced, in general, to the Toda or Volterra equations by Miura-type transformations. Moreover, equation~(\ref{Req}) is also related to the Landau--Lifshitz equation \cite{sy90}. A generalization of equation~(\ref{Req}) with nine arbitrary constant coef\/f\/icients has been considered in \cite{ly}.

By a straightforward computation we get the following result: all three-point generalized symmetries in the $n$-direction with no explicit dependence on~$n$,~$m$ for the ABS equations are particular cases of the YdKN equation. For the various equations of the ABS classif\/ication the coef\/f\/icients $c_i$, $1 \leq i \leq 6$, read
\begin{gather*}
\begin{array}{lll@{\,\,\,}l@{\,\,\,\,}l@{\,\,\,\,}ll}
 {\rm{{H1:}}} &   c_1=0 , &  c_2=0, &  c_3=0 , &  c_4=0,  &  c_5 =0,  \qquad \ \  c_6=1,   \\
 {\rm{{H2:}}} & c_1=0 , &  c_2=0, &  c_3=0 , &  c_4=0,  &  c_5 =1,   \qquad \ \ c_6=2 \alpha ,   \\
 {\rm{{H3:}}} &  c_1=0 , &  c_2=0, &  c_3=0 , &  c_4=1,  &  c_5 =0,  \qquad \ \ c_6= 2 \alpha \delta ,   \\
 {\rm{{Q1:}}}&  c_1=0 , &  c_2=0, &  c_3=-1 , &  c_4=1,  &  c_5 =0,  \qquad  \ \ c_6= \alpha^2 \delta^2,   \\
 {\rm{{Q2:}}}&   c_1=0 , &  c_2=0, &  c_3=1 , &  c_4=-1,  &  c_5 = -\alpha^2,  \ \ \ \,\, c_6=\alpha^4,    \\
 {\rm{{Q3:}}} &  c_1=0 , &  c_2=0, &  c_3= - 4 \alpha^2 , &  c_4=2 \alpha (\alpha^2+1),  &  c_5 =0,  \qquad \ \ c_6=- (\alpha^2-1)^2 \delta^2,    \\
 {\rm{{Q4:}}}&  c_1=1 , &  c_2= -\alpha, &  c_3=\alpha^2 , &  c_4=\frac{g_2}4 - \alpha^2,  &
 c_5 =\frac{\alpha g_2}4 {+} \frac{g_3}2,  \, c_6= \frac{g_2^2}{16} + \alpha g_3.
\end{array}\!\!
\end{gather*}

\begin{proposition} \label{pp2}
The ABS equations {\rm (H1)--(H3)} and {\rm (Q1)--(Q4)} correspond to B\"acklund
transformations of the particular cases of the YdKN equation \eqref{Req}
listed above. The same holds for the non-autonomous ABS equations, such that
$\alpha$ is constant and $\beta = \beta_m$ or $\alpha = \alpha_n$ and $\beta$
is constant. Equation~\eqref{Req} and the replacement $u_i \rightarrow u_{i,0}$
provide the three-point generalized symmetries in the $n$-direction of the ABS
equations with  a constant $\alpha$ and $\beta = \beta_m$, while
equation~\eqref{Req} and the replacement $u_i \rightarrow u_{0,i}$, $\alpha
\rightarrow \beta$ provide symmetries in the $m$-direction for the case
$\alpha = \alpha_n$ and a constant $\beta$.
\end{proposition}

The non-autonomous case is brief\/ly discussed in \cite{ra} where they state that if $\alpha$ is not constant, then the ABS equations have no local three-point symmetries in the $n$-direction. We shall present three-, f\/ive- and many-point generalized symmetries in the $m$-direction
for such equations in Subsection \ref{MS}.

A relation between the ABS equations and dif\/ferential-dif\/ference equations is discussed \linebreak \mbox{in~\cite{abs,asu}}. In \cite{abs} most of the ABS equations are interpreted as nonlinear superposition princip\-les for dif\/ferential-dif\/ference equations of the form
\begin{gather}
\left(\partial_x u_{n+1} \right) \left(\partial_x u_{n}\right) = h(u_{n+1}, u_n; \alpha) ,  \label{BeTr}
\end{gather}
where $h$ is a polynomial of $u_{n+1}$, $u_n$. Equations of the form (\ref{BeTr}) def\/ine B\"acklund transformations for subcases of the Krichever--Novikov equation
\begin{gather}
\partial _t u = \partial_{xxx}u - \frac32 \frac{(\partial_{xx}u)^2 - P(u)}{\partial_x u } ,   \label{KN}
\end{gather}
where $P$ is a fourth degree polynomial with arbitrary constant coef\/f\/icients. In the case of equations~(H1) and (H3) with $\delta=0$, the corresponding dif\/ferential-dif\/ference equations have a~dif\/ferent form, and the resulting KdV-type equations dif\/fer from equation~(\ref{KN}).

In \cite{asu} it is shown  that the continuous limit of equation~(Q4) goes into a subcase of the YdKN equation. It is stated that equation~(Q4) def\/ines a B\"acklund transformation for the same subcase. The same scheme holds for equations~(Q1)--(Q3), but it is not clear if the resulting Volterra-type equations are of the form~(\ref{Req}).

\subsection{Miura transformations revised}

It is possible to revise the Miura transformations constructed in Section \ref{sec3} from the point of view of the generalized symmetries.

Let us introduce the following function
\begin{gather}
r_0=r(u_0,u_{-1})= A_0 u_{-1}^2 + 2 B_0 u_{-1} + C_0= R(u_{-1}, u_0, u_{-1}).
\nonumber \end{gather}
It can be checked that $r(u_0, u_{-1}) = r(u_{-1}, u_0)$ and, in terms of $r_0$ the right hand side of  equation~(\ref{Req}) reads
\begin{gather}
\frac{R_0}{u_1 - u_{-1}} = \frac{r_0}{u_1 - u_{-1}} + \frac12 \partial_{u_{-1}}
{r_0}= \frac {r_1}{u_1 - u_{-1}} - \frac12 \partial_{u_{1}}
r_1 . \label{iden}
\end{gather}

All the ABS equations, up to equation~(Q4), are such that $c_1 = c_2 = 0$, so that the polyno\-mial~$R_0$ is of second degree. In this case equation~(\ref{Req}) may be transformed \cite{y06} into equation~(\ref{vtype}) with $  f(u_{1}, u_0, u_{-1})= u_0 (u_{1} - u_{-1})$ (Volterra equation) by the Miura transformation
\begin{gather}
\tilde u_0 = - \frac {r_1}{(u_2 - u_0) (u_1 - u_{-1})} .
\nonumber \end{gather}

The above map brings any solution $u_0$ of equation~(\ref{Req}) with $c_1 = c_2 = 0$ into a solution $\tilde u_0$ of the Volterra equation. This is exactly the same Miura transformation we have already presented in Section~\ref{sec3}. So, also at the level of the generalized symmetries, we may  see that there is a deep relation between equations~(H1)--(H3) and (Q1)--(Q3) and the Volterra equation.
If equation~(\ref{Req}) cannot be transformed to the case with $c_1 = c_2 = 0$, using a M\"obius  transformation, then it cannot be mapped into the Volterra equation by $\tilde u_0 = G(u_0, u_1, u_{-1}, u_2, u_{-2}, \dots)$~\cite{y06}. Equation~(Q4) is of this kind and thus is the only equation of the ABS list which cannot be related to the Volterra equation.

\subsection{Master symmetries}\label{MS}

Generalized symmetries of equation~(\ref{Req}) will also  be  compatible with the ABS equations, which
are, according to Proposition \ref{pp2}, their B\"acklund transformations. Such symmetries can be constructed, using the master symmetry of equation~(\ref{Req}) presented in
\cite{asy}.

Let us rewrite equation~(\ref{Req}) by using the equivalent $n$-dependent notation
(see equation~(\ref{not})), namely
\begin{gather}
 \frac{d u_{n}}{d \ep_0}= f_n^{(0)}= \frac {R(u_{n+1}, u_n, u_{n-1})}{u_{n+1} - u_{n-1}}
 \label{Reqn},
\end{gather}
where $\ep_0$ is the continuous symmetry parameter (previously
denoted with $\ep$). We shall denote with $\ep_i$, $i \geq 1$, the parameters
corresponding to higher generalized symmetries
\begin{gather*}
\frac{d u_{n}}{d \ep_i}= f_n^{(i)}, \qquad {\rm such \  that} \qquad
\frac{d f_n^{(j)}}{d \ep_i}  - \frac{d  f_n^{(i)}}{d \ep_j} =0,
\qquad i,j \geq 0.
\end{gather*}

Let us  introduce the master symmetry
\begin{gather}
\frac{d u_{n}}{d \tau}=g_n , \qquad {\rm such \  that} \qquad
\frac{d f_n^{(i)}}{d \tau}- \frac{d g_n}{d \ep_i}   =f_n^{(i+1)}, \qquad i \geq 0. \label{constr}
\end{gather}
Once we know the master  symmetry (\ref{constr})
we can construct explicitly the inf\/inite
hierarchy of generalized
symmetries.

The master symmetry of equation~(\ref{Reqn}) is given by
\begin{gather}
g_n = n f_n^{(0)}.   \label{ms}
\end{gather}
According to a general procedure described in \cite{y06} we need to introduce an explicit dependence on the parameter $\tau$
into the master symmetry (\ref{ms}) and into equation~(\ref{Reqn}) itself.
Let the coef\/f\/icients~$c_i$, appearing in the polynomials
$A_n$, $B_n$, $C_n$, be functions of
$\tau$. This $\tau$-dependence implies that~$r_n$ satisf\/ies the following partial dif\/ferential equation
\begin{gather}
2 \partial_\tau r_n = r_n \partial_{u_n} \partial_{u_{n-1}} r_n- \left(\partial_{u_n} r_n \right) \left(\partial_{u_{n-1}}
r_n \right) .  \label{rtau}
\end{gather}
On the left hand side of the above equation, we dif\/ferentiate only the
coef\/f\/icients of $r_n$ with respect to $\tau$. The right hand side has the same
form as $r_n$, but with dif\/ferent coef\/f\/icients. Collecting the
coef\/f\/icients of the  terms $u_n^i u_{n-1}^j$ for various powers $i$ and $j$, we obtain a system of six
ordinary dif\/ferential equations for the six coef\/f\/icients $c_i(\tau)$, whose initial conditions are $c_i(0)=c_i$.
Generalized symmetries constructed
by using equation~(\ref{constr}) explicitly depend on $\tau$. They remain generalized
symmetries for any value of $\tau$, as $\tau$ is just a parameter for them and for equation~(\ref{Reqn}). So, going over to the initial conditions, we get generalized
symmetries of  equation~(\ref{Reqn}) and of the corresponding ABS equations.

Let us derive, as an illustrative example, a formula for the
symmetry $f_n^{(1)}$ from equation~(\ref{constr}).
From equations~(\ref{Reqn})--(\ref{ms}) it follows that
\begin{gather}
f_n^{(1)} = \partial_\tau f_n^{(0)} + f_{n+1}^{(0)} \partial_{u_{n+1}}
f_n^{(0)} - f_{n-1}^{(0)} \partial_{u_{n-1}} f_n^{(0)}.
\label{f1}
\end{gather}
Using equations~(\ref{iden}) and (\ref{rtau}) one obtains
\begin{gather}
\partial_{u_{n+1}} f_n^{(0)} = - \frac {r_n}{(u_{n+1} - u_{n-1})^2} ,
\qquad \partial_{u_{n-1}} f_n^{(0)} = \frac {r_{n+1}}{(u_{n+1} - u_{n-1})^2} ,
\nonumber \end{gather}
and
\begin{gather}
\partial_\tau R_n =
\CR_n= \CR (u_{n+1},u_n ,u_{n-1})
= \CA_n \,   u_{n+1} u_{n-1} +
\frac{\CB_n}{2} (u_{n+1}+ u_{n-1}) + \CC_n ,
\nonumber \end{gather}
with
\begin{gather*}
\CA_n =B_n \partial_{u_n} A_n   -A_n \partial_{u_n} B_n , \qquad
\CB_n = C_n\partial_{u_n}  A_n  - A_n\partial_{u_n} C_n,  \\
\CC_n = C_n \partial_{u_n}  B_n  - B_n\partial_{u_n}  C_n.
\end{gather*}

From equation~(\ref{f1}) we get the f\/irst generalized symmetry
\begin{gather}
\frac{d u_{n}}{d \ep_1}= f_n^{(1)} = \frac {\CR_n}{u_{n+1} - u_{n-1}} - \frac {r_n f_{n+1}^{(0)}  + r_{n+1}
f_{n-1}^{(0)}  }{(u_{n+1} - u_{n-1})^2} .  \label{gs1}
\end{gather}
Up to our knowledge this formula is new. It provides f\/ive-point generalized symmetries in both $n$- and $m$-directions for the ABS equations. Examples of such   f\/ive-point symmetries for equations~(H1) and (Q1) with $\delta = 0$ can be found in \cite{ra,gr1}.

Let us clarify the construction of the symmetry $f_n^{(1)}$ for
equations~{\rm{(H1)--(H3)}}. In these cases the function $r_n$ takes the form
\begin{gather}
r_n = 2 c_4(\tau) u_n u_{n-1} + 2 c_5(\tau) (u_n + u_{n-1}) + c_6(\tau),
\nonumber \end{gather}
and equation~(\ref{rtau}) is equivalent to the system
\begin{gather}
\partial_\tau c_4(\tau) = 0 , \qquad \partial_\tau c_5(\tau)  = 0 , \qquad
 \partial_\tau c_6(\tau)= c_4(\tau) c_6(\tau) - 2 c^2_5(\tau) .  \label{syst}
\end{gather}
The initial conditions of system (\ref{syst}) are (see the list above Proposition \ref{pp2})
\begin{alignat*}{5}
& {\rm{{H1:}}} \quad && c_4(0)=0, \qquad &&  c_5(0)=0, \qquad && c_6(0)=1, & \\
& {\rm{{H2:}}} \quad && c_4(0)=0, \qquad && c_5(0)=1, \qquad && c_6(0)=2 \alpha , & \\
& {\rm{{H3:}}} \quad && c_4(0)=1, \qquad && c_5(0)=0, \qquad && c_6(0)=2 \alpha \delta, &
\end{alignat*}
and its solutions are given by
\begin{alignat*}{5}
& {\rm{{H1:}}} \quad && c_4(\tau)=0, \qquad && c_5(\tau)=0, \qquad && c_6(\tau)=1, & \\
& {\rm{{H2:}}} \quad && c_4(\tau)=0, \qquad && c_5(\tau)=1, \qquad && c_6(\tau)=2( \alpha -\tau), & \\
& {\rm{{H3:}}} \quad && c_4(\tau)=1, \qquad && c_5(\tau)=0, \qquad && c_6(\tau)=2 \alpha \delta e^\tau. &
\end{alignat*}

Note that the master symmetry with the above $c_i(\tau)$ generates $\tau$-dependent symmetries for a~$\tau$-dependent equation,
but by f\/ixing $\tau$ we obtain $\tau$-independent symmetries for a
$\tau$-independent equation. Let us remark that the $\tau$-dependence
is independent of the order of the symmetry and it may be used for the
construction of all higher symmetries.

So, according to formula (\ref{gs1}), we may construct the
generalized symmetry $f_n^{(1)}$,
in the case of the list H, from the following expressions
\begin{alignat*}{5}
& {\rm{{H1:}}} \quad && f_n^{(0)} =\frac{1}{u_{n+1}-u_{n-1}} , \qquad &&  r_n =1 , \qquad &&   \CR_n=0, & \\
& {\rm{{H2:}}} \quad && f_n^{(0)}= \frac{u_{n+1}+u_{n-1} + 2 (u_n+ \alpha)}{u_{n+1}-u_{n-1}} ,  \qquad && r_n = 2 (u_n+ u_{n-1}+ \alpha),
\qquad &&  \CR_n= -2 , & \\
& {\rm{{H3:}}} \quad && f_n^{(0)}= \frac{u_n (u_{n+1}+u_{n-1}) + 2 \alpha \delta}{u_{n+1}-u_{n-1}}, \qquad && r_n = 2 ( u_n u_{n-1} + \alpha \delta),
\qquad && \CR_n = 2 \alpha \delta . &
\end{alignat*}
It is possible to verify that the symmetries (\ref{gs1})  with
$f_n^{(0)}$, $r_n$, $\CR_n$ given above are compatible with
both equations~(\ref{Reqn}) and (H1)--(H3).

By using the master symmetry constructed above we can construct  inf\/inite hierarchies of many-point
generalized symmetries of the ABS equations in both directions. In the non-autonomous cases
(see Proposition \ref{pp2}) we provide one hierarchy in the $n$- or $m$-direction.
The master symmetry and formula (\ref{gs1}) will also be useful in the case of
the generalizations of the ABS equations presented in the next Subsection. It
should be remarked that in \cite{ra} the authors constructed  master
symmetries for all autonomous and non-autonomous ABS equations, which are of a
dif\/ferent kind with respect to the ones presented here.

%%%%%%%%%%%%%%%%%%%%%%%%%%%%%%%
\subsection{Generalizations of the ABS equations} \label{sec5last}
%%%%%%%%%%%%%%%%%%%%%%%%%%%%%%%
Here we discuss the generalization of the ABS equations introduced by Tongas,
Tsoubelis and Xenitidis (TTX) in~\cite{gr}. The TTX equations are autonomous
lattice equations of the form~(\ref{jj}) which possess only two of the four
main properties of the ABS equations: they are af\/f\/ine linear and possess the symmetries of the square.

In terms of the polynomial $\mathcal E$, see equation~(\ref{jj}), one generates the following function $h$
\[
h(u_{0,0}, u_{1,0}; \alpha, \beta) = \mathcal E \partial_{u_{0,1}} \partial_{u_{1,1}}
\mathcal E-
\left(  \partial_{u_{0,1}} \mathcal E \right) \left(   \partial_{u_{1,1}} \mathcal E  \right),
\]
which is a biquadratic and symmetric polynomial in its f\/irst two arguments. It
has been proved in \cite{gr} that the TTX equations admit three-point
generalized symmetries in the $n$-direction of the form
\begin{gather}
\frac{d u_{0,0}}{d \ep} = \frac {h}{u_{1,0} - u_{-1,0}} - \frac12 \partial_{u_{1,0}}
h  .  \label{grsym}
\end{gather}
Of course, there is a similar symmetry in the $m$-direction. Comparing equations~(\ref{Req}), (\ref{iden}) and~(\ref{grsym}), we see that the symmetry (\ref{grsym}) is nothing but the YdKN equation in its general form.
This shows that all TTX equations can also be considered as B\"acklund transformations for the YdKN equation. However, they probably describe the general picture for B\"acklund transformations of the YdKN equation, which have the form~(\ref{jj}). The general formula (\ref{gs1}) and the master symmetry discussed in the previous Subsection, provide f\/ive- and many-point generalized symmetries of the TTX equations in both directions, thus
conf\/irming  their integrability.

\section{Concluding remarks} \label{sec5}

In this paper we have considered some further properties of the ABS
equations. In particular we have shown that equations (H1)--(H3) and
(Q1)--(Q3) can be transformed into equations associated with the
spectral problem of the Volterra equation. Therefore  all known results for
the solution of the Volterra equation can be used to construct solutions of
the ABS equations.
Moreover, all equations of the ABS list, except equation~(Q4), can be transformed
among
themselves by Miura transformations.

The situation of equation~(Q4) is somehow dif\/ferent. It
is shown that this equation can be thought as a B\"acklund transformation
for a subcase of the Yamilov discretization of the Krichever--Novikov
equation. But it cannot be related by a Miura transformation to a
Volterra-type equation and this explains the complicate form of its scalar
spectral problem. The master symmetry constructed for the YdKN equation can,
however, be used also in this case to construct generalized symmetries.

It turns out that a generalizations of the ABS equations introduced by Tongas, Tsoubelis and Xenitidis are B\"acklund transformations for the YdKN equation.

Further generalizations  of the TTX and ABS equations can be probably obtained by a proper  explicit dependence on the point of the lattice not only in the lattice parameters $\alpha$ and $\beta$, but also in the $\mathbb{Z}^2$-lattice equation itself.
The existence of an $n$-dependent generalization of the YdKN equation,
introduced in~\cite{ly},  could help in solving this problem. Such a generalization
is integrable in the sense that it has a master symmetry~\cite{asy} similar to
the one presented here.

\subsection*{Acknowledgments}

DL, MP and CS have been partially supported by PRIN Project {\it Metodi geometrici nella
teoria delle onde non lineari ed applicazioni-2006} of the Italian Minister for Education
and Scientif\/ic Research. RY has been partially supported by the Russian Foundation for Basic Research (Grant numbers 07-01-00081-a and 06-01-92051-KE-a) and he thanks the University of Roma Tre for hospitality. This work has been done in the framework of the Project {\it Classification of integrable discrete and continuous models} f\/inanced by a joint grant from EINSTEIN consortium and RFBR.

\pdfbookmark[1]{References}{ref}
\LastPageEnding


\begin{thebibliography}{99}

\footnotesize\itemsep=-0.5pt

\bibitem{a2000}
Adler V.E.,
On the structure of the B\"acklund transformations for the relativistic lattices,
{\it J. Nonlinear Math. Phys.} {\bf 7} (2000), 34--56, \href{http://arxiv.org/abs/nlin.SI/0001072}{nlin.SI/0001072}.

\bibitem{abs}
Adler V.E., Bobenko A.I., Suris Yu.B.,
Classif\/ication of integrable equations on quad-graphs. The consistency approach,
{\it Comm. Math. Phys.} {\bf 233} (2003), 513--543, \href{http://arxiv.org/abs/nlin.SI/0202024}{nlin.SI/0202024}.

\bibitem{abs1}
Adler V.E., Bobenko A.I., Suris Yu.B.,
Discrete nonlinear hyperbolic equations. Classif\/ication of integrable cases,
\href{http://arxiv.org/abs/0705.1663}{arXiv:0705.1663}.

\bibitem{asy}
Adler V.E., Shabat A.B., Yamilov R.I.,
The symmetry approach to the integrability problem,
{\it Teoret. Mat. Fiz.} {\bf 125} (2000), 355--424 (English transl.: {\it Theoret. and Math. Phys.} {\bf125} (2000), 1603--1661).

\bibitem{asu}
Adler V.E., Suris Yu.B.,
${\rm Q}_4$: integrable master equation related to an elliptic curve,
{\it Int. Math. Res. Not.} {\bf 2004} (2004), no.~47, 2523--2553, \href{http://arxiv.org/abs/nlin.SI/0309030}{nlin.SI/0309030}.

\bibitem{AV}
Adler V.E., Veselov A.P.,
Cauchy problem for integrable discrete equations on quad-graph,
{\it Acta Appl. Math.} {\bf 84} (2004), 237--262, \href{http://arxiv.org/abs/math-ph/0211054}{math-ph/0211054}.

\bibitem{atkinson}
Atkinson J.,
B\"acklund transformations for integrable lattice equations,
{\it J. Phys. A: Math. Theor.} {\bf 41} (2008) 135202, 8~pages, \href{http://arxiv.org/abs/0801.1998}{arXiv:0801.1998}.


\bibitem{atk2}
Atkinson J., Hietarinta J., Nijhof\/f F.W.,
Seed and soliton solutions for Adler's lattice equation,
{\it J. Phys. A: Math. Theor.} {\bf 40} (2007), F1--F8, \href{http://arxiv.org/abs/nlin.SI/0609044}{nlin.SI/0609044}.

\bibitem{bob}
Bobenko A.I., Suris Yu.B.,
Integrable systems on quad-graphs,
{\it Int. Math. Res. Not.} {\bf 2002} (2002), no.~11, 573--611, \href{http://arxiv.org/abs/nlin.SI/0110004}{nlin.SI/0110004}.

\bibitem{ck}
Case K.M., Kac M.,
A discrete version of the inverse scattering problem,
{\it J. Math. Phys.} {\bf 14} (1973), 594--603.

\bibitem{cl}
Chiu S.C., Ladik J.F.,
Generating exactly soluble nonlinear discrete evolution equations
by a generalized Wronskian technique,
{\it J. Math. Phys.} {\bf 18} (1977), 690--700.

\bibitem{emp}
Francoise J.P., Naber G., Tsou S.T. (Editors), Encyclopedia of mathematical physics,
 Elsevier, 2007.

\bibitem{galor}
Galor O.,
Discrete dynamical systems, Springer, Berlin, 2007.

\bibitem{hi}
Hirota R.,
Nonlinear partial dif\/ference equations. I.~A~dif\/ference analog of the
Korteweg--de Vries equation,
{\it J. Phys. Soc. Japan} {\bf 43} (1977), 1423--1433.\\
Hirota R.,
Nonlinear partial dif\/ference equations. III. Discrete sine-Gordon equation, {\it J. Phys. Soc. Japan} {\bf 43} (1977), 2079--2086.

\bibitem{kn}
Krichever I.M., Novikov S.P.,
Holomorphic bundles over algebraic curves, and nonlinear equations,
{\it Uspekhi Mat. Nauk} {\bf 35} (1980), no.~6, 47--68 (in Russian).

\bibitem{l}
Levi D.,
Nonlinear dif\/ferential-dif\/ference equations as B\"acklund transformations,
{\it J. Phys. A: Math. Gen.} {\bf 14} (1981), 1083--1098.



\bibitem{lp}
Levi D., Petrera M.,
Continuous symmetries of the lattice potential KdV equation,
{\it J. Phys. A: Math. Theor.} {\bf 40} (2007), 4141--4159, \href{http://arxiv.org/abs/math-ph/0701079}{math-ph/0701079}.

\bibitem{lps}
Levi D., Petrera M., Scimiterna C.,
The lattice Schwarzian KdV equation and its symmetries,
{\it J. Phys. A: Math. Theor.} {\bf 40} (2007), 12753--12761, \href{http://arxiv.org/abs/math-ph/0701044}{math-ph/0701044}.

\bibitem{lw6}
Levi D., Winternitz P.,
Continuous symmetries of dif\/ference equations,
{\it J. Phys. A: Math. Gen.} {\bf  39} (2006),  R1--R63, \href{http://arxiv.org/abs/nlin.SI/0502004}{nlin.SI/0502004}.

\bibitem{ly}
Levi D., Yamilov R.I.,
Conditions for the existence of higher symmetries of evolutionary equations on the lattice,
{\it J. Math. Phys.} {\bf 38} (1997), 6648--6674.

\bibitem{msy}
Mikhailov A.V., Shabat A.B., Yamilov R.I.,
The symmetry approach to the classif\/ication of nonlinear equations. Complete lists of integrable systems,
{\it Uspekhi Mat. Nauk} {\bf 42} (1887), no.~4, 3--53 (English transl.: {\it Russian Math. Surveys} {\bf 42} (1987), no.~4, 1--63).


\bibitem{ni}
Nijhof\/f F.W.,
Lax pair for the Adler (lattice Krichever--Novikov) system,
{\it Phys. Lett. A} {\bf 297} (2002), 49--58, \href{http://arxiv.org/abs/nlin.SI/0110027}{nlin.SI/0110027}.


\bibitem{nc}
Nijhof\/f F.W., Capel H.W.,
The discrete Korteweg--de Vries equation,
{\it Acta Appl. Math.} {\bf 39} (1995), 133--158.

\bibitem{ra}
Rasin O.G., Hydon P.E.,
Symmetries of integrable dif\/ference equations on the quad-graph,
{\it Stud. Appl. Math.} {\bf 119} (2007), 253--269.

\bibitem{sande}
Sandevan J.T.,
Discrete dynamical systems. Theory and applications, The Clarendon Press,
Oxford University Press,
New York, 1990.

\bibitem{sy90}
Shabat A.B., Yamilov R.I.,
Symmetries of nonlinear chains,
{\it Algebra i Analiz} {\bf 2} (1990), 183--208 (English transl.: {\it Leningrad Math. J.} {\bf 2} (1991), 377--400).

\bibitem{gr1}
Tongas A., Tsoubelis D., Papageorgiou V.,
Symmetries and group invariant reductions of integrable partial dif\/ference equations,
in Proc. 10th Int. Conf. in Modern Group Analysis (October 24--31, 2004, Larnaca, Cyprus),
Editors N.H.~Ibragimov, C.~Sophocleous and P.A.~Damianou, 2004, 222--230.

\bibitem{gr}
Tongas A., Tsoubelis D., Xenitidis P.,
Af\/f\/ine linear and $D_4$ symmetric lattice equations: symmetry analysis and reductions,
{\it J. Phys. A: Math. Theor.} {\bf 40} (2007), 13353--13384, \href{http://arxiv.org/abs/0707.3730}{arXiv:0707.3730}.


\bibitem{y94}
Yamilov R.I., Construction scheme for discrete Miura transformations,
{\it J. Phys. A: Math. Gen.} {\bf 27} (1994), 6839--6851.

\bibitem{y83}
Yamilov R.I., Classif\/ication of discrete evolution equations,
{\it Uspekhi Mat. Nauk} {\bf 38} (1983), no.~6, 155--156 (in Russian).

\bibitem{y06}
Yamilov R.I., Symmetries as integrability criteria for dif\/ferential dif\/ference equations,
{\it J. Phys. A: Math. Gen.} {\bf 39} (2006), R541--R623.

\end{thebibliography}
\end{document}